\documentclass[10pt,twocolumn]{article} 
\usepackage{simpleConference}
\usepackage{times}
\usepackage{graphicx}
\usepackage{amssymb}
\usepackage{url,hyperref}
\usepackage{lmodern}
\usepackage{rotating}
\usepackage{xcolor}
\usepackage{latexsym}
\usepackage{amsmath}
\usepackage{bbm}
\usepackage{booktabs}
\usepackage{tabularx} 
\usepackage{array}

\begin{document}

\title{Metrics that Matter: Evaluating Image Quality Metrics for Medical Image Generation}

\author{
Yash Deo\textsuperscript{1,*}, Yan Jia\textsuperscript{1}, Toni Lassila\textsuperscript{2}, William A.~P.~Smith\textsuperscript{1}, Tom Lawton\textsuperscript{1,8}, \\
Siyuan Kang\textsuperscript{3}, Alejandro F. Frangi\textsuperscript{4,5,6,7}, Ibrahim Habli\textsuperscript{1} \\
\\
\textsuperscript{1}Department of Computer Science, University of York, York, UK \\
\textsuperscript{2}School of Computer Science, University of Leeds, Leeds, UK \\
\textsuperscript{3}Department of Computing and Mathematics, \\ Manchester Metropolitan University, Manchester, UK \\
\textsuperscript{4}Department of Computer Science, School of Engineering, University of Manchester, \\  Manchester, UK \\
\textsuperscript{5}School of Health Sciences, Division of Informatics, Imaging and Data Sciences,\\ University of Manchester, Manchester, UK \\
\textsuperscript{6}Department of Cardiovascular Sciences, KU Leuven, Leuven, Belgium \\
\textsuperscript{7}Department of Electrical Engineering (ESAT), KU Leuven, Leuven, Belgium \\
\textsuperscript{8}Improvement Academy, Bradford Institute for Health Research, Bradford Royal Infirmary,\\ Bradford, UK \\
}
\maketitle
\thispagestyle{empty}

\begin{abstract}
Evaluating generative models for synthetic medical imaging is crucial yet challenging, especially given the high standards of fidelity, anatomical accuracy, and safety required for clinical applications. Standard evaluation of generated images often relies on no-reference image quality metrics when ground truth images are unavailable, but their reliability in this complex domain is not well established. This study comprehensively assesses commonly used no-reference image quality metrics using brain MRI data, including tumour and vascular images, providing a representative exemplar for the field. We systematically evaluate metric sensitivity to a range of challenges, including noise, distribution shifts, and, critically, localised morphological alterations designed to mimic clinically relevant inaccuracies. We then compare these metric scores against model performance on a relevant downstream segmentation task, analysing results across both controlled image perturbations and outputs from different generative model architectures. Our findings reveal significant limitations: many widely-used no-reference image quality metrics correlate poorly with downstream task suitability and exhibit a profound insensitivity to localised anatomical details crucial for clinical validity. Furthermore, these metrics can yield misleading scores regarding distribution shifts, e.g. data memorisation. This reveals the risk of misjudging model readiness, potentially leading to the deployment of flawed tools that could compromise patient safety. We conclude that ensuring generative models are truly fit for clinical purpose requires a multifaceted validation framework, integrating performance on relevant downstream tasks with the cautious interpretation of carefully selected no-reference image quality metrics. 
\end{abstract}

\section{Introduction}
\label{sec:introduction}

The development of advanced medical imaging applications, particularly those leveraging artificial intelligence (AI), often requires large, diverse, and accurately labelled datasets. However, acquiring such datasets faces significant hurdles due to patient privacy regulations and the high cost and complexity associated with expert anatomical annotation. These limitations can constrain the training and validation of potentially beneficial AI tools. Generative deep learning models offer a potential solution to these challenges. Techniques such as generative adversarial networks (GANs, see \cite{singh2021} for a review), variational autoencoders (VAEs), and denoising diffusion probabilistic models (DDPMs, see \cite{kazerouni2023} for a review) have demonstrated the capacity to synthesise (or generate, terms used synonymously herein) high-fidelity medical images. These models encompass various approaches, including conditional generation (e.g., image translation \cite{li2024fddmunsupervisedmedicalimage}) and unconditional synthesis, with generated images being used for tasks including data augmentation (\cite{kebaili2023}), privacy preservation (\cite{liu2024}), cross-modality synthesis (\cite{fard2022}), image super-resolution (\cite{ahmad2022}), and virtual imaging trials (\cite{abadi2020}). Regardless of the specific generation strategy, robust evaluation of the output's quality and reliability is essential, particularly distinguishing the nature of these synthetic outputs from direct empirical measurements~\cite{Parker2021_Virtually}.

While promising, the integration of generative models and the synthetic data they produce into clinical research and potentially downstream applications necessitates rigorous validation to ensure they are trustworthy and safe. Unlike directly measured medical images, outputs from generative models function more like complex computational simulations or predictions derived from learned distributions~\cite{Roush2018, Parker2022_Evidence}. Evaluating these typically requires \emph{no-reference image quality metrics} (NRIQMs), as a direct reference often doesn't exist. These metrics must assess not only image \emph{fidelity} (resemblance to real images) but also \emph{diversity} (adequate coverage of the true data distribution)~\cite{Autptch}, moving beyond simple similarity to evaluate the model's representation of reality~\cite{Leonelli2019}. Generative models can suffer from failure modes such as \emph{mode collapse} (overproducing similar images from certain modes with limited diversity across the data distribution) or \emph{mode invention} (creating spurious, 'hallucinated' images with features not present in the real data). The deployment of clinical decision support tools trained or validated using synthetic data from poorly evaluated generative models poses significant risks to patient safety. For instance, mode invention could lead to downstream tools learning from hallucinated pathological features, while mode collapse might yield tools biased against rare conditions; without the right risk controls or guardrails, reliance on such compromised tools could subsequently lead to inaccurate diagnoses or inappropriate treatment selections. Therefore, the thorough evaluation of the image generation process using appropriate and validated metrics, considering their 'adequacy-for-purpose'~\cite{BokulichParker2021, Parker2023_Eval}, is not merely a technical requirement but a crucial step for ensuring the safety and reliability of clinical applications ultimately impacted by synthetic data.

In this work, we perform a systematic assessment of commonly used NRIQMs intended for evaluating generative medical imaging outputs. We subject these NRIQMs to rigorous experiments designed to stress-test both their sensitivity to common image artifacts (e.g., noise) and their ability to detect known generative model failure modes (e.g., mode collapse, mode invention, data memorisation, morphological inaccuracies). We compare results from \emph{upstream} evaluation, where quantitative NRIQMs are applied directly to the synthetic images to assess their intrinsic properties and distributional characteristics, against performance in \emph{downstream} evaluation. For the purpose of this study, downstream evaluation specifically involves assessing the utility of the generated images by inputting them into a relevant, pre-trained task model – in this case, a segmentation model – and measuring the accuracy or quality of that model's output when processing the synthetic input. This comparative approach allows us to investigate which upstream metrics robustly reflect this defined task-specific utility or detect critical image flaws, and which provide potentially misleading evaluations of model performance. We focus our validation on brain magnetic resonance imaging (MRI), utilising brain tumour images from the BRaTS dataset \cite{brats1,brats2,brats3} and vascular angiography images from the IXI dataset \cite{IXI}. Although our study uses brain MRI, the methodology and findings regarding metric behaviours are potentially extensible to other medical imaging modalities. Our implementation is available for adaptation/use at the following GitHub repository: https://github.com/YashDeo-York/GenMed.

Our primary contribution is a critical assessment of NRIQMs for generative models in medical imaging. Specifically, we provide concrete evidence showing that popular distance metrics (e.g. FID, KID, MMD) can yield misleadingly optimistic scores under conditions such as data memorisation or mode collapse. Critically, our work highlights a profound limitation across nearly all tested upstream metrics: their insensitivity to localised, clinically relevant morphological alterations, such as distorted tumour boundaries, questioning their ability to ensure anatomical accuracy. Furthermore, we demonstrate a significant and recurring discrepancy where upstream metric rankings of generative models (VAE, GAN, DDPM) do not align with their performance on downstream segmentation tasks. Our analysis therefore underscores the urgent need for evaluation metrics specifically validated for medical imaging and provides strong evidence supporting the necessity of multifaceted evaluation frameworks that prioritise assessments of 'adequacy-for-purpose'~\cite{BokulichParker2021, Parker2023_Eval} to ensure generative models, and the applications built upon them, meet the stringent safety and reliability standards required in clinical contexts.

\section{Background and Related Work}

The evaluation of image-generative models requires a comprehensive approach that balances structural fidelity, perceptual realism, and task-specific performance. Generating medical images further demands metrics capable of capturing nuanced clinical and anatomical features.

Traditionally, for image quality evaluation, \emph{full-reference} (pixel-level) metrics, such as SSIM (structural similarity index measure) and PSNR (peak signal-to-noise ratio), are widely adopted for their simplicity and ability to quantify fine-grained reconstruction accuracy. However, these metrics require a reference image to compare against, which is challenging when no ground-truth image exists. Furthermore, full-reference metrics often fail to adequately account for perceptual quality or alignment with real-world data distributions.

To address these limitations, recent research has shifted towards \emph{NRIQMs} that evaluate image quality without requiring pixel-wise alignment to a ground truth. Among the most widely used are the Fréchet Inception Distance (FID)~\cite{FID} and the Inception Score (IS)~\cite{IS}. FID compares the statistical distributions of real and generated image features extracted from a pre-trained network (typically InceptionV3), providing a global measure of distributional similarity. IS, on the other hand, assesses both the quality and diversity of generated images by computing the entropy of predicted label distributions from an Inception classifier.

More recent alternatives, such as Kernel Inception Distance (KID)~\cite{KID}, use polynomial kernel-based Maximum Mean Discrepancy (MMD) and provide unbiased estimates that are better suited for small datasets. In parallel, several domain-agnostic metrics have emerged to capture specific generative failure modes. The \textbf{AuthPct}~\cite{Autptch} (Authenticity Percentage) measures the proportion of generated samples falling outside the support of the real distribution, helping to identify out-of-distribution generations and address issues related to mode invention. The \textbf{CT Score}~\cite{Ct} (Cluster Tightness Score) quantifies the compactness of feature clusters in the generated data, potentially revealing mode collapse or insufficient diversity generations. \textbf{FD$_\infty$}~\cite{FDinf}, an extension of Fréchet distance, captures worst-case divergence and highlights rare but significant deviations in feature space.

Despite the recent surge in new metrics proposed for evaluating generative model quality and distributional characteristics, including those designed to address mode collapse and mode invention, a significant gap exists in their validation on medical imaging data. Many metrics developed on natural image datasets do not adequately address the unique complexities of medical images, such as subtle pathological variations, noise sensitivity, and domain shifts arising from diverse scanner configurations. While specialised metrics tailored to medical imaging have been proposed (\cite{stein2024exposing}), their adoption remains limited.

\begin{figure}[!t]
  \centerline{\includegraphics[width=\columnwidth]{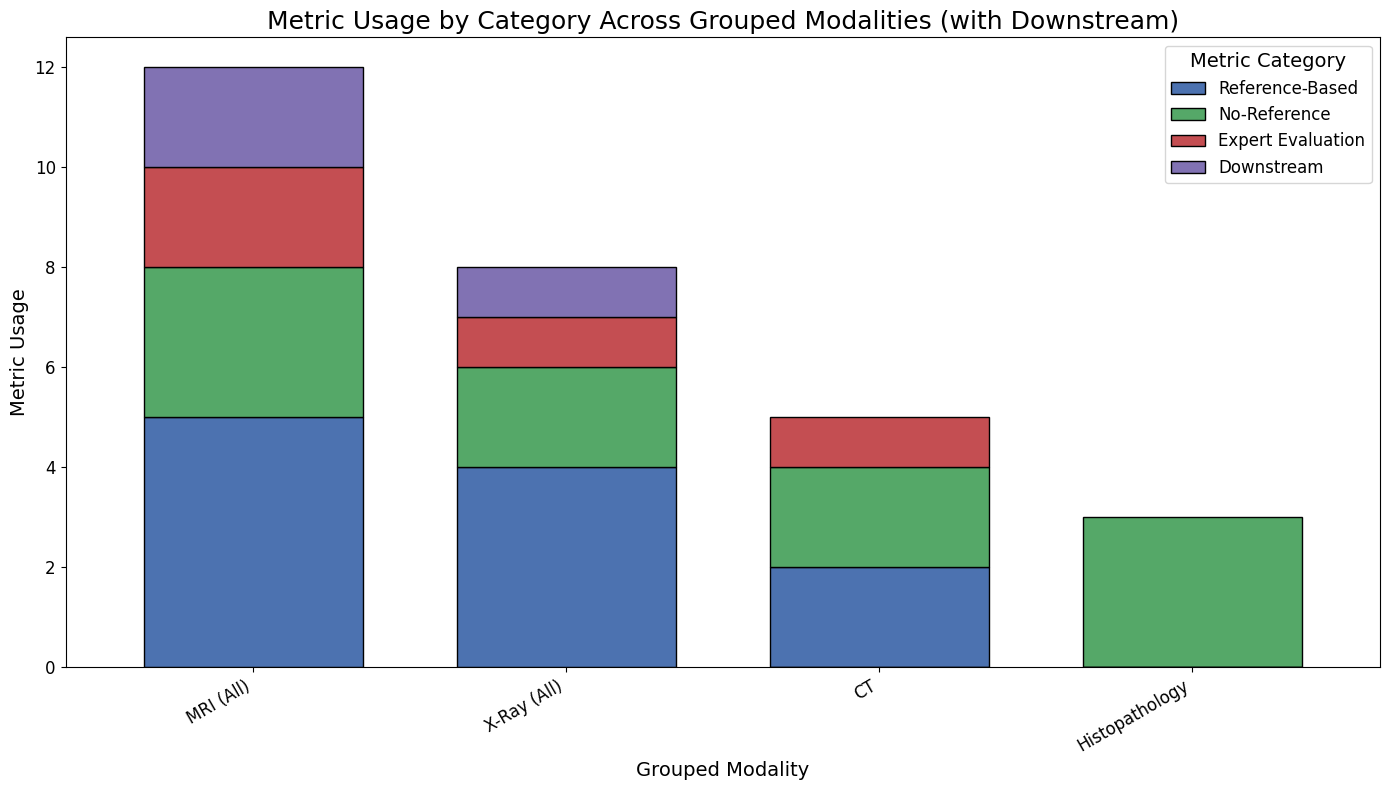}}
  \caption{A graph of different types of metrics used in generative medical image evaluation, categorised by modality.}
  \label{fig1}
\end{figure}

To explore current evaluation practices in the field, we conducted a literature review of papers employing generative models in medical imaging. Our search spanned publications from 2019 to 2024 across major databases including IEEE Xplore, PubMed, arXiv, and Google Scholar, using keywords operation \{"medical image synthesis," or "medical imaging,"\} and \{"GAN," or "diffusion models" or "VAE" or "Variational auto-encoder"\} and \{"reference-metrics" or "no-reference metrics" or "image quality assessment"\}. After filtering for relevance , We identified 28 papers. Through this review, we identify key trends in the usage of metrics for medical image generation in research, which we present in Fig~\ref{fig1}. 

Ideally, expert evaluation by clinicians should serve as the gold standard for assessing the anatomical plausibility and clinical utility of synthetic medical images. However, this approach faces significant practical challenges, including considerable time and cost requirements, inherent inter-observer variability where different experts may provide differing assessments, difficulties in scaling such evaluations to large datasets or numerous model iterations, and the need for carefully standardised rating protocols to ensure consistency. Despite these limitations, expert evaluation remains valuable and has been employed in specific contexts. For instance, ~\cite{clin1} trained diffusion models on CT scans of different organs and relied on radiologists to rank images by realism, inter-slice consistency, and anatomical correctness. Similarly, ~\cite{clin2} and ~\cite{clin3} applied GAN-based approaches to mammograms and brain MRI respectively, and asked clinicians to judge whether each image was real or fake. Other works have tried to establish correlations between expert evaluations and image quality metrics (~\cite{chow2016b}; ~\cite{mason2019})

In practice, as shown in Fig \ref{fig1}, the most commonly used metrics for evaluating synthetic medical images are full-reference metrics, particularly PSNR and various forms of SSIM, including MS-SSIM (Multi-scale SSIM)~\cite{MSSSIM} and 4GR-SSIM~\cite{4GRSSIM}). These metrics are especially common in settings where generative models are applied to tasks like image super-resolution~\cite{sr1,sr2}, slice synthesis or interpolation~\cite{s1,s2}, and image reconstruction~\cite{pix1,pix3,pix4,pix5,pix6}, where partial or proxy ground-truth images are available. Although these applications are not strictly paired in a supervised sense, they offer 'semi-ground-truth' data that makes using full-reference metrics for comparisons feasible. 

Interestingly, even in tasks involving unconditional generation — where no explicit reference image exists — researchers continue to use full-reference metrics. A common workaround involves identifying the closest real image to each generated image using nearest-neighbour search in pixel or feature space, followed by averaging metric scores over the dataset~\cite{deo2023shapeguidedconditionallatentdiffusion, deo2024few, cheng2024synthesising}. Other studies use MS-SSIM or 4GR-SSIM not to evaluate fidelity, but as a proxy for diversity~\cite{pix2,pix3,fet1}. The underlying assumption is that MS-SSIM values close to 1 suggest excessive similarity (potential mode collapse), while values too close to 0 indicate low-quality or noisy generations. While these techniques offer creative adaptations of full-reference metrics, they remain heuristic and lack rigorous theoretical justification.

Because full-reference metrics are inherently limited in their ability to capture perceptual quality and global distributional alignment between real and generated images, there has been a growing shift towards NRIQMs such as FID and IS~\cite{fet1,fet3,fet2}, which is the second most used metric category as shown in Fig \ref{fig1}. These metrics evaluate image quality based on high-dimensional features extracted from deep networks, offering a more holistic assessment of realism and diversity in unpaired or distribution-based generative tasks. However, the reliability of most NRQIMs is heavily influenced by the choice of feature extractor. Most implementations rely on features derived from networks pre-trained on natural image datasets, typically using InceptionV3. Due to concerns about the suitability of such networks for medical imaging — where visual patterns differ significantly from natural images — alternatives such as Med3D~\cite{MED3D} have been proposed. Med3D, pre-trained on large-scale medical imaging datasets, has recently been adopted in place of Inception-based extractors in several studies~\cite{deo2023shapeguidedconditionallatentdiffusion,fet1}. Conversely, subsequent research~\cite{gm1} has indicated that feature extractors specifically trained on medical datasets, including Med3D, may produce evaluation scores that are unstable or inconsistent when assessing generative outputs. Contrarily, the InceptionV3 network, despite its lack of domain-specific training, has demonstrated more stable results. Recently, academic focus has been directed towards self-supervised architectures such as DINOv2~\cite{stein2024exposing}, which possess the capability to learn broad visual representations. While these models have exhibited potential in natural image benchmarks, their efficacy and consistency within the domain of medical imaging remain substantially unvalidated.

Taken together, these contradictory findings highlight a broader uncertainty in the field: there is currently no standardised or well-validated framework for evaluating generative medical images using NRIQMs. The choice of feature extractor has a substantial impact on metric behaviour, yet the absence of consensus and the limited validation of these methods in clinical contexts continue to hinder the development of consistent and interpretable evaluation pipelines.

Alongside these developments, there is a growing trend towards using downstream tasks as a means of evaluating the quality and clinical utility of synthetic images, accounting for a proportion similar to that of expert evaluation in our literature review, as illustrated in Fig \ref{fig1}. This approach is particularly valuable when generative models are developed for task-specific purposes, such as simulating anatomical variations or pathological features. For example, in the context of brain vasculature generation, \cite{deo2023shapeguidedconditionallatentdiffusion} proposed training a classifier to predict the anatomical type of generated vessels as a proxy for realism. Similarly, \cite{deo2024few} suggested that performing downstream flow simulations on generated aneurysm-containing vessels could serve as a strong indicator of structural validity. In tumour imaging, \cite{V1} evaluated the realism of synthetic brain MRI images by passing them through a pre-trained tumour segmentation network, while \cite{V2} assessed segmentation accuracy by comparing predicted label maps from generated images against reference maps.

While downstream evaluation offers a promising framework for assessing practical utility, it does not provide a complete picture of generative performance. Specifically, downstream tasks are often insensitive to properties such as sample diversity, distributional coverage, and visual fidelity — dimensions that are critical in understanding model behaviour. Furthermore, the absence of ground truth for generated images introduces uncertainty when interpreting downstream performance. Although recent work has proposed ground-truth-free metrics to assess segmentation accuracy on generated images~\cite{senbi2024groundtruthfreeevaluationsegmentationmedical}, these approaches are still in early stages and have not yet seen widespread adoption.

Therefore, there is a clear need to systematically investigate the sensitivity and reliability of different evaluation metrics for medical image generation — both from upstream and downstream perspectives. While some prior work has attempted to assess the performance of image quality metrics in medical imaging, the scope has been limited. For instance, \cite{dohmen2025} provided a review of image quality metrics for image-to-image translation tasks but did not include NRIQMs. Other studies~\cite{chow2016,kastryulin2023,leveque2021} have examined medical image quality metrics more broadly, yet did not focus on generative image evaluation.

In this work, we adopt a structured and objective approach to evaluate NRIQMs in a clinically relevant setting. We first conduct a series of controlled perturbation experiments — including noise injection, morphological manipulations, and domain shifts — on real data to assess how different metrics respond to well-characterised variations. This helps to quantify each metric's sensitivity to common failure modes in generative image analysis. Subsequently, we evaluate the outputs of multiple generative models (VAE, GAN, DDPM) using the same metrics, and further assess their utility through downstream clinical tasks. This allows us to examine how well metric responses align with real-world use cases. Together, this framework offers a more nuanced understanding of the strengths and limitations of NRIQMs in the context of medical image generation.
 
\subsection{Metrics used for Upstream Evaluation}
\label{sec:metrics}

This section summarises NRIQMs used in this study to assess synthetic medical images where a ground truth reference image is typically unavailable. We divide the selected metrics into three complementary categories based on the primary aspect of quality they aim to capture: \textit{distribution alignment metrics}, \textit{perceptual realism metrics}, and \textit{statistical fidelity metrics}. A concise technical summary, including formulas and references, is provided in Table~\ref{tab:metrics_summary_revised}. 

\paragraph{(A) Distribution Alignment Metrics}
These metrics quantify how well the distribution of generated image features aligns with that of real data features, primarily evaluating whether a model captures the overall diversity and statistical moments of the true distribution.

\begin{itemize}
  \item \textbf{FID (Fr\'echet Inception Distance)}: Computes the Fr\'echet distance between Gaussian distributions fitted to feature embeddings (typically from InceptionV3) of real versus generated images, comparing their means and covariances. Lower FID values indicate closer distribution alignment. While sensitive to various artifacts, the choice of feature space can impact results; consistent use of standard InceptionV3 features is common for comparability (\cite{FID}). We follow recent findings suggesting the reliability of Inception features even for medical images~\cite{gm1}.

  \item \textbf{KID (Kernel Inception Distance)}: Similar to FID but uses polynomial kernel comparisons (specifically, the squared Maximum Mean Discrepancy, MMD) between feature distributions, offering potentially greater stability for smaller sample sizes often encountered in medical datasets (\cite{KID}).

  \item \textbf{sFID (Spatial FID)}: Extends standard FID by incorporating spatial information from feature maps, aiming to provide sensitivity to structural misalignments (e.g., organ shape, lesion location) which are critical in medical imaging (\cite{SFID}).

  \item \textbf{PRDC (Precision, Recall, Density, Coverage)}: A suite of metrics evaluating distribution similarity based on feature manifolds. \emph{Precision} measures the fraction of generated samples falling within the real data manifold (fidelity), while \emph{Recall} measures the fraction of the real data manifold covered by generated samples (diversity). High recall is crucial for capturing rare conditions, while high precision helps avoid unrealistic artifacts (\cite{PRDC,IPR}). Density and Coverage provide further insights into relative densities and manifold coverage.

  \item \textbf{Vendi Score}: Measures diversity based on the similarity between samples within a set, calculated using a kernel function over feature vectors. Higher scores indicate greater diversity. It aims to capture the spread of the distribution (\cite{vendi}).
\end{itemize}

\paragraph{(B) Perceptual Realism Metrics}
These metrics focus on qualities related to human perception, structural plausibility, or distinguishability from real images, often assessing individual samples or local features.

\begin{itemize}
  \item \textbf{LPIPS (Learned Perceptual Image Patch Similarity)}: Primarily a reference-based metric correlating well with human perception by evaluating patch-wise similarity between two images in deep feature space~\cite{LPIPS}. However, it can be adapted for evaluating generative models in a no-reference manner by analysing aggregate perceptual distances. Common adaptations involve calculating the average pairwise LPIPS distance among generated samples (where lower average distances may indicate lower diversity or mode collapse) or computing the average LPIPS distance between each generated image and its closest neighbour in the real dataset (assessing perceptual fidelity to the real data manifold). 

  \item \textbf{Inception Score (IS)}: Assesses image quality (sharpness/distinctiveness) and diversity using predicted class probabilities from a pre-trained classifier (Inception). Higher scores suggest high-quality, diverse images. Its applicability to medical imaging can be limited by the need for relevant, well-defined class labels (\cite{IS}).

  \item \textbf{Realism Score}: Often employs a discriminator or classifier trained to distinguish real from generated images. High scores (or low discriminator accuracy) suggest generated samples are visually and feature-wise difficult to distinguish from real images (\cite{Realism}). 
\end{itemize}

\paragraph{(C) Statistical Fidelity \& Manifold Metrics} 
These metrics target statistical properties of the generated distribution, often focusing on how well generated samples adhere to the underlying manifold of real data in feature space, complementing global distribution alignment measures.

\begin{itemize}
  \item \textbf{FLS (Feature Likelihood Score)}: Quantifies the likelihood of generated features under a distribution model (often Gaussian) fitted to real data features. Higher FLS values suggest synthetic images are statistically typical of the real data manifold (\cite{FLS}).

  \item \textbf{ASW (Approximate Sliced Wasserstein)}: Approximates the Wasserstein distance between real and generated image feature distributions using random projections (slices). Lower values indicate closer distributions. It can be interpreted in terms of feature space clustering (\cite{ASW}).

  \item \textbf{Ct Score (Clusterability Score)}: Evaluates distribution overlap by assessing how well generated image features conform to clusters established by real image features in feature space. A lower Ct Score indicates better adherence to real data clusters (\cite{Ct}). 

  \item \textbf{AuthPct (Authenticity Percentage)}: Calculates the fraction of generated samples falling within a region defined by the real data distribution in feature space, based on a learned threshold. It measures the proportion of synthetic images deemed statistically plausible or \emph{authentic} (\cite{Autptch}).

  \item \textbf{FD$_\infty$ (Infinite Fr\'echet Distance)}: Measures the maximum divergence between feature distributions along any projection, capturing worst-case discrepancies rather than average differences, as in FID. Potentially sensitive to outliers or rare failure modes (\cite{FDinf}).
\end{itemize}

\medskip
\noindent

Table~\ref{tab:metrics_summary_revised} consolidates the key formulas and parameter definitions for each metric mentioned above.

\begin{table*}[htpb]
\centering
\caption{Summary of Metrics Used in the Evaluation}
\label{tab:metrics_summary_revised} 
\begin{tabular}{|l|p{7cm}|p{7cm}|} 
\hline
\textbf{Name} & \textbf{Description} & \textbf{Formula / Computation Principles} \\ 
\hline
FID & Fréchet Inception Distance: Evaluates similarity between real and generated feature distributions using mean and covariance of features (typically InceptionV3). Lower is better. & $\text{FID} = ||\mu_r - \mu_g||_2^2 + \text{Tr}(\Sigma_r + \Sigma_g - 2(\Sigma_r\Sigma_g)^{1/2})$ \\ 
\hline
Spatial FID (sFID) & Spatially-aware FID extension incorporating structural context by using features from intermediate convolutional layers. Lower is better. & Computed similarly to FID, but using spatially-preserved features. \\ 
\hline
Inception Score (IS) & Assesses quality (low conditional entropy) and diversity (high marginal entropy) using an Inception classifier's predictions on generated images. Higher is better. & $\text{IS} = \exp\left( \mathbb{E}_x D_{KL}(p(y|x) || p(y)) \right)$ \\ 
\hline
KID & Kernel Inception Distance: Compares feature distributions using the squared Maximum Mean Discrepancy (MMD) with a polynomial kernel. Lower is better. & $\text{KID} = \text{MMD}^2(\mathcal{X}_r, \mathcal{X}_g)$ computed on feature sets $\mathcal{X}_r, \mathcal{X}_g$. \\ 
\hline
LPIPS & Learned Perceptual Image Patch Similarity: Reference-based metric comparing patch-wise deep features; lower distance implies higher perceptual similarity. & $\text{LPIPS}(x_0, x_1) = \sum_l w_l ||\Phi_l(x_0) - \Phi_l(x_1)||_2^2$ over layers $l$. \\ 
\hline
ASW & Approximate Sliced Wasserstein Distance: Measures Wasserstein-1 distance between feature distributions using random 1D projections ($\theta$). Lower is better. & $\text{ASW} = \mathbb{E}_{\theta \sim S^{d-1}} W_1(P_\theta(\mathcal{X}_r), P_\theta(\mathcal{X}_g))$ \\ 
\hline
FLS & Feature Likelihood Score: Quantifies how statistically typical generated features are under a model (e.g., Gaussian) fitted to real features. Higher is better. & Typically involves fitting $p_{\text{real}}(f)$ to real features $f_r$ and evaluating $\mathbb{E}_{f_g \sim P_g} [\log p_{\text{real}}(f_g)]$. (Note: Original formula matched FID, likely incorrect for likelihood). \\ 
\hline
DreamSIM & Learned perceptual similarity metric trained to align with human judgments, comparing deep features. Higher is better (assuming similarity). & $\text{DreamSIM}(x_0, x_1) = f(\Phi(x_0), \Phi(x_1))$, where $f$ is a learned similarity function. \\ 
\hline
FD$_\infty$ & Infinite Fréchet Distance (often related to MMD): Measures worst-case discrepancy between distributions via maximizing difference over functions in a unit ball. Lower is better. & $\text{FD}_\infty \approx \sup_{f \in \mathcal{F}} \left| \mathbb{E}_{x \sim P_r}[f(x)] - \mathbb{E}_{x \sim P_g}[f(x)] \right|$ for function class $\mathcal{F}$. \\ 
\hline
Ct Score & Clusterability Score: Evaluates how well generated features fit within clusters defined by real features. Lower score implies better fit/less dispersion. & Conceptually measures properties like variance or distance within/between clusters formed by real vs. generated features. (Note: Original formula seemed incomplete/unclear). \\ 
\hline
AuthPct & Authenticity Percentage: Fraction of generated samples considered `authentic' or statistically plausible relative to the real data distribution. Lower indicates more divergence. & $\text{AuthPct} = \frac{1}{N_g} \sum_{i=1}^{N_g} \mathbbm{1}_{\{x_{g,i} \in \mathcal{D}_{\text{real}}\}}$. (Note: $\mathcal{D}_{\text{real}}$ typically defined via threshold $\tau$ on density $p(x)$ or distance $d(x, \text{real data})$). \\ 
\hline
PRDC & Precision, Recall, Density, Coverage for Distributions: Evaluates sample quality (Precision) and distribution coverage (Recall) based on nearest-neighbour analysis in feature space. Higher is better. & Precision $\approx$ Fraction of generated samples whose NN is real. Recall $\approx$ Fraction of real samples whose NN is generated. (Simplified description). \\ 
\hline
Vendi Score & Measures diversity within a set of samples based on feature similarity, calculated using a kernel matrix over features. Higher score means higher diversity. & Computation involves eigenvalue decomposition of the kernel matrix $K_{ij} = k(\Phi(x_i), \Phi(x_j))$. \\ 
\hline
Realism Score & Measures the probability that a classifier/discriminator judges a generated image as real.Higher is better. & $\text{Realism}(x_g) = P(\text{real} \mid x_g)$ using a trained discriminator/classifier. \\ 
\hline
\end{tabular}
\end{table*}

\subsection{Downstream Evaluation Technique}
\label{subsec:downstream technique}
To assess the practical utility and clinical relevance of the generated images beyond upstream quantitative metrics, we evaluated their suitability for a critical downstream task: automated segmentation. We employed relevant, pre-trained segmentation algorithms appropriate for each dataset: brain vessel segmentation for the synthetic IXI MRA images~\cite{deo2023learnedlocalattentionmaps} and brain tumour segmentation for the synthetic BraTS MRI images~\cite{feng}. These algorithms were applied directly to the synthetic images obtained from the pre-trained VAE, GAN, and DDPM models.

\subsubsection{Addressing the Reference Image Challenge in Downstream Evaluation} %
\label{subsubsec:ref_challenge}

A fundamental challenge in evaluating downstream tasks on purely synthetic images generated from noise or latent variables is the absence of corresponding ground truth annotations (e.g., segmentation masks) for comparison. Direct pixel-level metrics such as Dice or IoU cannot be computed without a reference. While qualitative visual assessment of the segmentation outputs on synthetic images is possible, it introduces subjectivity. To overcome this reference-free challenge and enable quantitative assessment, we adopt the regression-based segmentation quality evaluation methodology proposed by Senbi et al.~\cite{senbi2024groundtruthfreeevaluationsegmentationmedical}.

This involves training an auxiliary regression model—the quality prediction model—to estimate segmentation quality scores (specifically, Dice scores in our case) based on the input image and a generated segmentation mask, without requiring the ground truth mask during the final evaluation phase. Our implementation involved the following key steps:

\begin{enumerate}
\item \textbf{Target Segmentation Generation:} First, we obtained the segmentations whose quality needed to be assessed. This involved applying standard, pre-trained segmentation models (specific to each dataset, see~\cite{feng, deo2023learnedlocalattentionmaps}) to the input images (either the perturbed real images from controlled experiments or the synthetic images from generative models).
\item \textbf{Quality Predictor Training Data:} Second, we created a diverse dataset to train the quality prediction model itself. For each real image in our training set, we first generated an initial segmentation mask using the standard pre-trained segmentation model. We then created multiple synthetically perturbed versions of this mask using various augmentations (e.g., adding noise, simulating small erosions/dilations). Crucially, since we possessed the ground truth masks for these real images, we calculated the accurate Dice score between each original or perturbed mask and the corresponding ground truth mask. The training dataset for the quality predictor thus consisted of pairs of {real image, potentially perturbed segmentation mask}, with its precisely calculated Dice score serving as the target regression label. This process generated multiple training samples per original image, covering a spectrum of segmentation qualities. (Note: Alternative training data strategies mentioned in the literature, such as using only pairs of {real image, real ground truth mask} assuming a perfect score, were not the primary method employed here).
\item \textbf{Quality Predictor Training:} Third, we trained the regression model (A ResNet classifier) to predict the Dice score based on the input image and its corresponding segmentation mask from the training data created in Step 2.
\item \textbf{Downstream Quality Assessment Step:} Finally, this trained quality prediction model (from Step 3) was used in both experimental phases to assess downstream quality. It was applied to pairs of {input image (perturbed real or synthetic), corresponding segmentation mask (generated in Step 1)} to provide the predicted Dice score (referred to as the Evanyseg score in our results), estimating the quality of the segmentation achievable on each input image without direct access to ground truth.
\end{enumerate}

We acknowledge that evaluating segmentation quality using such a predictive model introduces its own layer of potential uncertainty – the evaluation relies on the accuracy of the trained quality predictor. However, in the absence of ground truth for generated images, this method provides a standardised, quantitative, and reproducible proxy for downstream task performance. It allows for a relative comparison of how well the different generative models produce images suitable for the pre-trained segmentation task, moving beyond purely subjective visual inspection or solely relying on upstream NRIQMs. While not a perfect substitute for ground-truth validation, it serves as a necessary tool for assessing downstream task suitability in this challenging reference-free context.

\section{Methodology}
\label{sec:method}

\begin{figure*}

\centerline{\includegraphics[width=\textwidth]{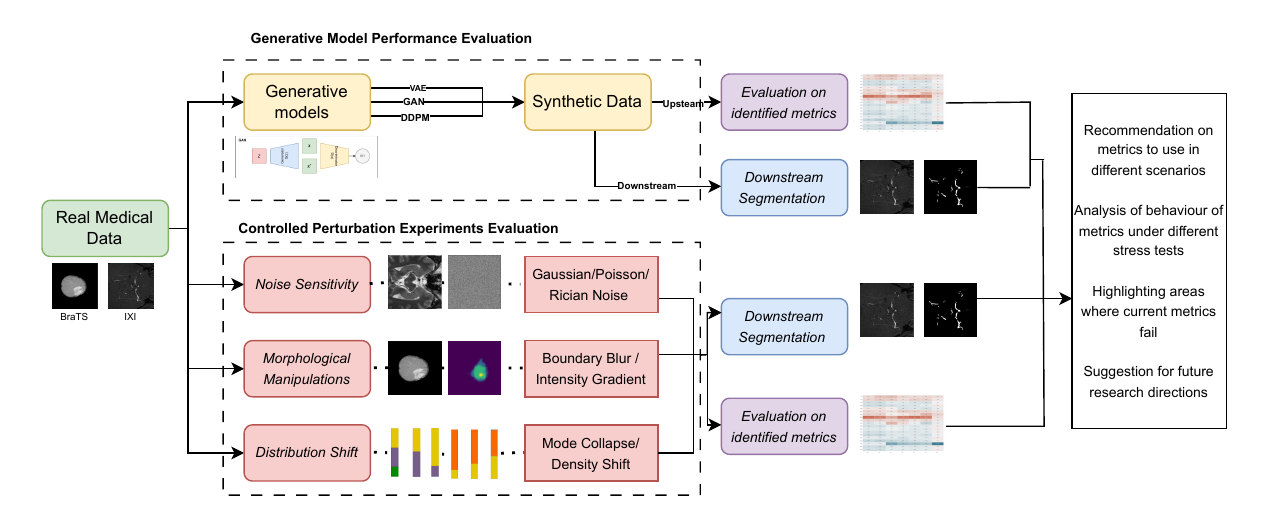}}
\caption{Overview of our methodology. }
\label{overall}
\end{figure*}

To systematically evaluate the suitability of NRIQMs for generative medical imaging, our methodology employs a unified evaluation framework integrating both upstream quantitative metric assessment and downstream task validation across two distinct phases. Phase one involves \emph{controlled perturbation experiments evaluation} where known alterations are introduced to real medical images (MRI and MRA) to simulate specific artifacts or distributional issues. Phase two focuses on \emph{generative model performance evaluation}, assessing synthetic images produced by representative generative architectures (VAE, GAN, DDPM). Crucially, within \textbf{both} phases, we apply the full suite of selected upstream NRIQMs directly to the images (perturbed real images in Phase 1 and synthetic images in Phase 2) \textit{and} perform downstream evaluations by assessing the performance of a pre-trained segmentation model when applied to these same images. This comprehensive approach allows us to analyse metric sensitivity under controlled conditions, assess performance on actual generative model outputs, and critically compare whether upstream metric scores align with downstream task utility across diverse scenarios. By integrating insights from both phases, we aim to provide robust evidence regarding the strengths and limitations of common NRIQMs for evaluating generative models in clinically relevant contexts. Fig.~\ref{overall} showcases a high-level overview of this integrated methodology.

\subsection{Phase 1: Controlled Perturbation Experiments Evaluation}
\label{sec:phase1_controlled_experiments} 

To systematically evaluate NRIQM robustness under specific conditions, this phase involves applying controlled perturbations to real medical images. We utilise subsets of real images from the BRaTS MRI (tumour-focused) and IXI MRA (vessel-focused) datasets as our \emph{base image sets}. We then create multiple distinct \emph{perturbed image sets} by applying controlled manipulations to copies of these base images. These manipulations, detailed in the following sections, simulate specific challenges including noise injection, morphological alterations, and various forms of distribution shift (e.g., data memorisation, mode changes). For metrics requiring comparison to a reference distribution (such as FID, KID, PRDC, etc.), we utilise a separate, unperturbed set of real images from the corresponding dataset, designated as the \emph{reference distribution set}. This experimental design allows us to isolate and analyse the response of both upstream NRIQMs and downstream task performance (using the pre-trained segmentation model) to specific, well-defined image perturbations across different anatomical structures and imaging characteristics.

\section*{\textbf{Noise Sensitivity}}
\label{sec:noise_exp}

Noise is an intrinsic characteristic of medical imaging, arising from scanner hardware limitations, acquisition protocols, and patient movement (in the form of motion artefacts). Generative models must learn to either faithfully replicate these noise patterns or appropriately compensate for them. However, different architectures might introduce distinct noise signatures, and evaluation metrics may respond inconsistently to scanner-induced versus model-generated noise. We therefore conduct structured noise-injection experiments using three clinically relevant noise models (Gaussian, Poisson, Rician) applied to our base image sets, allowing us to assess both upstream metric responses and downstream task performance under these conditions.

\paragraph{\bf Gaussian Noise Injection}
Probabilistic generative models, particularly VAEs and some diffusion models, frequently introduce Gaussian-like noise artifacts or smoothing in their reconstructions due to their underlying mathematical formulation~\cite{VAE}. Since VAEs often optimise a likelihood function based on a Gaussian prior, their outputs can exhibit characteristic smoothness with subtle noise textures. Similarly, diffusion models may retain residual Gaussian-like noise depending on their noise schedule and de-noising process. This experiment assesses how effectively evaluation metrics and downstream tasks capture the impact of these specific noise patterns.
Experimental Design: We introduce zero-mean Gaussian noise with progressively increasing standard deviations ($\sigma=\{0.01, 0.05, 0.1\}$) to copies of the real MRA base images. These noise levels are calibrated to simulate gradual degradation. We then compute the suite of upstream NRIQMs and evaluate downstream segmentation performance on perturbed image sets. We assess the degree to which metrics and task performance change relative to the noise level, identifying potential insensitivities to this type of degradation.

\paragraph{\bf Rician and Poisson Noise Injection}
Unlike Gaussian noise, magnitude MRI acquisitions typically contain Rician-distributed noise resulting from the reconstruction process of complex-valued k-space data~\cite{gudbjartsson1995rician}. Poisson noise represents an intensity-dependent noise process relevant to certain low-count imaging scenarios. Generative models often struggle to accurately reproduce these specific noise distributions, sometimes over-denoising images (removing clinically relevant texture) or introducing non-physical artifacts. This experiment evaluates whether metrics and downstream tasks are sensitive to degradations reflecting more realistic noise characteristics encountered in MRI.
Experimental Design: We apply calibrated levels of Rician and Poisson noise to copies of the MRA base images. For Rician noise, we implement the standard model:
\[
I_{\text{rician}} = \sqrt{(I_{\text{orig}} + n_1)^2 + n_2^2}
\]
where $n_1, n_2 \sim \mathcal{N}(0, \sigma^2)$ represent independent Gaussian noise components. For Poisson noise, intensity variations are scaled appropriately for the image data. We systematically vary the noise level (via $\sigma$ or equivalent SNR) to create a spectrum of conditions. For each noise type and level, we compute the full suite of upstream NRIQMs and assess downstream segmentation performance on perturbed image sets.

\paragraph{Metric and Task Response Analysis} 
For each noise type, we analyse the response curves of the upstream metrics (e.g., via z-score of percentage change from baseline) and the corresponding changes in downstream segmentation performance scores. We quantify the sensitivity of both metrics and the downstream task to each noise type, observing the rate of change per unit increase in noise level. Comparing sensitivities across noise types (e.g., high sensitivity to Gaussian vs. Rician for a given metric) provides insights into potential metric biases. A metric that demonstrates high sensitivity to Gaussian noise but remains relatively insensitive to Rician noise likely reflects optimisation for natural image domains rather than medical-specific degradation patterns. Conversely, metrics that appropriately penalise Rician and Poisson noise — which better represent real MRI acquisition physics — may be more suitable for evaluating medical image generation models. Critically, by comparing the magnitude of upstream metric responses to the degradation observed in downstream task performance, we can assess how well different metrics reflect the practical impact of noise on image utility for downstream tasks. These controlled noise-injection experiments establish quantitative baselines for understanding metric and task behaviour under well-characterised perturbations.

\section*{\textbf{Morphological Manipulations}}
\label{sec:morph_exp}

Morphological alterations in medical images, particularly within pathological regions like tumours, are often critical indicators for diagnosis, staging, and treatment planning. Generative models aiming to synthesise realistic medical images must accurately represent and preserve these structural details. Therefore, evaluation metrics should ideally be sensitive to even subtle variations in relevant anatomical or pathological morphology. We conduct controlled morphological alteration experiments, focusing specifically on tumour regions within the BRaTS MRI base images, using two distinct approaches: tumour boundary blurring and intensity-based tumour gradient variations. We assess the impact of these manipulations using both upstream NRIQMs and downstream segmentation performance.

\paragraph{\bf Tumour Boundary Blurring}
Accurate delineation of tumour boundaries is crucial for precise surgical planning and radiotherapy target definition. Generative models that produce overly smooth or indistinct boundaries can lead to clinically significant errors in downstream applications such as volume estimation or treatment planning~\cite{clark2006automatic}, compromising patient safety. This experiment simulates such scenarios and evaluates whether metrics or downstream tasks can detect these boundary inaccuracies.
Experimental Design: Using the ground truth segmentation masks available in the BRaTS dataset, we identify the tumour regions in the base MRI images. To selectively blur the boundaries, we first compute the Euclidean distance transform within the tumour mask to identify boundary regions based on proximity to the edge. A standard Gaussian blurring filter was then applied specifically to the intensity values within these identified boundary regions on each slice. The strength of the blurring effect was controlled by systematically adjusting the standard deviation of the Gaussian kernel across several distinct levels, simulating varying degrees of boundary sharpness produced by generative models. Subsequently, we compute the upstream NRIQMs and evaluate downstream tumour segmentation performance on these boundary-manipulated images.

\paragraph{\bf Intensity-Based Tumour Gradient Variations}
Beyond boundary shape, the internal intensity characteristics and texture gradients of tumours provide valuable diagnostic information, potentially indicating different tissue compositions or stages of development. Generative models should ideally reproduce these internal patterns accurately. Changes in tumour intensity can reflect variations in vascularity or necrosis, influencing treatment strategy. This experiment investigates whether metrics or downstream tasks can detect simulated changes in internal tumour intensity gradients.

\paragraph{\bf Experimental Design} Within the identified tumour regions of the base MRI images, we introduce artificial intensity gradients based on spatial location relative to the tumour's centre of mass (calculated per slice). Specifically, we applied a \emph{radial intensity gradient}, modulating pixel intensities multiplicatively based on their distance from the tumour centre using a sinusoidal function to create structured variations. The magnitude of this radial intensity variation was systematically adjusted across several levels to simulate different degrees of internal tumour heterogeneity that might be distorted or poorly represented by generative models. The upstream NRIQMs were then computed, and downstream tumour segmentation performance was evaluated on the perturbed image set with modified internal gradients to assess sensitivity.

\paragraph{\bf Metric and Task Response Analysis} 
For both tumour boundary blurring and intensity-based gradient variations, we analyse the response of the upstream evaluation metrics and the downstream segmentation task performance. We examine how metric scores and segmentation accuracy/quality change as the degree of morphological manipulation increases. Metrics or downstream task evaluations that show significant changes are deemed more sensitive to these clinically relevant structural variations. Conversely, metrics or downstream task evaluations that remain relatively stable despite these alterations may fail to capture these important details, highlighting potential blind spots in the evaluation process. Comparing the upstream metric responses (or lack thereof) to any observed changes in downstream performance is particularly crucial for understanding whether the metrics reflect the practical impact of these subtle morphological changes.
 
\section*{\textbf{Distribution Shift}}
\label{sec:domain_exp}

Distribution shifts pose a significant challenge in medical imaging, where generative models must accurately represent the inherent variability of clinical data without introducing biases or failing to capture the full spectrum of real-world cases. While metrics have been proposed to detect distribution shifts in natural(non-medical) image generation, their efficacy and interpretation in the context of subtle medical image variations require specific investigation. Such shifts can manifest in various ways, including data memorisation, mode collapse, density disparities due to different acquisition sources, and mode invention (hallucinations). We conduct a series of controlled experiments designed to evaluate the sensitivity of both upstream NRIQMs and downstream task performance to these specific types of distribution shifts.

\paragraph{\bf Data Memorisation}
Generative models, particularly high-capacity architectures like diffusion models, have been observed to memorise or closely replicate training data examples, which can lead to overly optimistic performance evaluations if not detected~\cite{akbar2025}. We assess metrics and downstream task sensitivity to two forms of data memorisation: external data duplication - replication of training examples and internal data duplication - duplication within the generated set itself.
Experimental Design: We perform these two types of duplication experiments using the base image sets. To simulate external duplication, representing training set replication leakage into a test set, we create perturbed sets by gradually replacing base images with exact copies of images assumed to belong to a hypothetical 'training' partition, using replacement rates of [5\%, 15\%, 30\%, 45\%]. To simulate internal duplication, reflecting a lack of diversity within a generated set, we create perturbed sets by introducing duplicates within the base image set at the same rates [5\%, 15\%, 30\%, 45\%]. For both scenarios, we then compute upstream NRIQMs and evaluate downstream segmentation performance to assess their sensitivity to these forms of data memorisation.

\paragraph{\bf Mode Collapse}
Mode collapse, a known issue particularly for some GAN architectures, occurs when a model fails to capture the full diversity of the data distribution, instead generating samples from only a limited subset of modes~\cite{modec1,modec2}. In medical imaging, this could mean failing to generate examples of rarer anatomies or pathologies. We investigate this by manipulating the known class distribution within the IXI MRA dataset, categorised by anatomical variations of the posterior communicating artery (PCoA) in the Circle of Willis (bilateral, unilateral, or absent).
Experimental Design: Our base MRA dataset has an approximate distribution of 40\% bilateral PCoA, 40\% unilateral PCoA, and 20\% absent PCoA anatomies. We create two perturbed image sets simulating mode collapse: (1) we entirely remove the 'absent PCoA' class (the rarest mode), resulting in a 50\%-50\% distribution between the remaining two classes; and (2) we further alter the distribution to an imbalanced 1:5 ratio between the bilateral and unilateral PCoA classes, still excluding the absent PCoA class. We analyse how both upstream NRIQMs and downstream vessel segmentation performance respond to these explicit shifts in class distribution, assessing their ability to detect the missing or under-represented modes.

\paragraph{\bf Density Shift}
 Medical imaging data is often aggregated from diverse sources (e.g., different scanners or sites), potentially leading to shifts in image intensity distributions or textures due to variations in hardware, software, or acquisition protocols. Metrics should ideally be robust to, or capable of detecting, such domain shifts. We simulate this using the IXI MRA dataset, which contains images acquired using two different Philips scanners (Intera 3T and Gyroscan Intera 1.5T) with slightly varying parameters (\cite{IXI}).
Experimental Design: We start with a base image set containing a balanced 50\%-50\% distribution of images from each scanner. We then create perturbed image sets with deliberately skewed distributions [10-90, 25-75, 75-25, 90-10] representing dominance of one scanner's data over the other. We evaluate how both upstream NRIQMs and downstream vessel segmentation performance respond to these density shifts arising from scanner differences.

\paragraph{\bf Mode Invention}
Mode invention occurs when a generative model produces outputs containing features or structures not present in the training data distribution, essentially creating unrealistic or spurious content (hallucinations). In medical imaging, this could manifest as anatomically implausible structures or tissue characteristics. Metrics should ideally penalise such inventions.
Experimental Design: We simulate subtle contrast-based mode invention by manipulating T2-weighted MRI base images from the IXI dataset. We apply controlled intensity remapping specifically within grey matter and white matter regions (identified using percentile-based intensity thresholds). A small, opposing scaling factor is applied to each tissue type, slightly increasing grey matter intensity while proportionally reducing white matter intensity, thus inverting their relative contrast locally in a physically implausible manner. This introduces novel structural appearances absent from the original data. Gaussian smoothing is applied post-scaling to maintain some anatomical coherence, followed by intensity normalisation. By repeating this manipulation with varying levels of contrast distortion, we evaluate whether upstream NRIQMs and downstream segmentation performance can detect these synthetic modes that deviate subtly but systematically from the real data distribution.

\begin{figure*}[!h]
\centerline{\includegraphics[width=\textwidth]{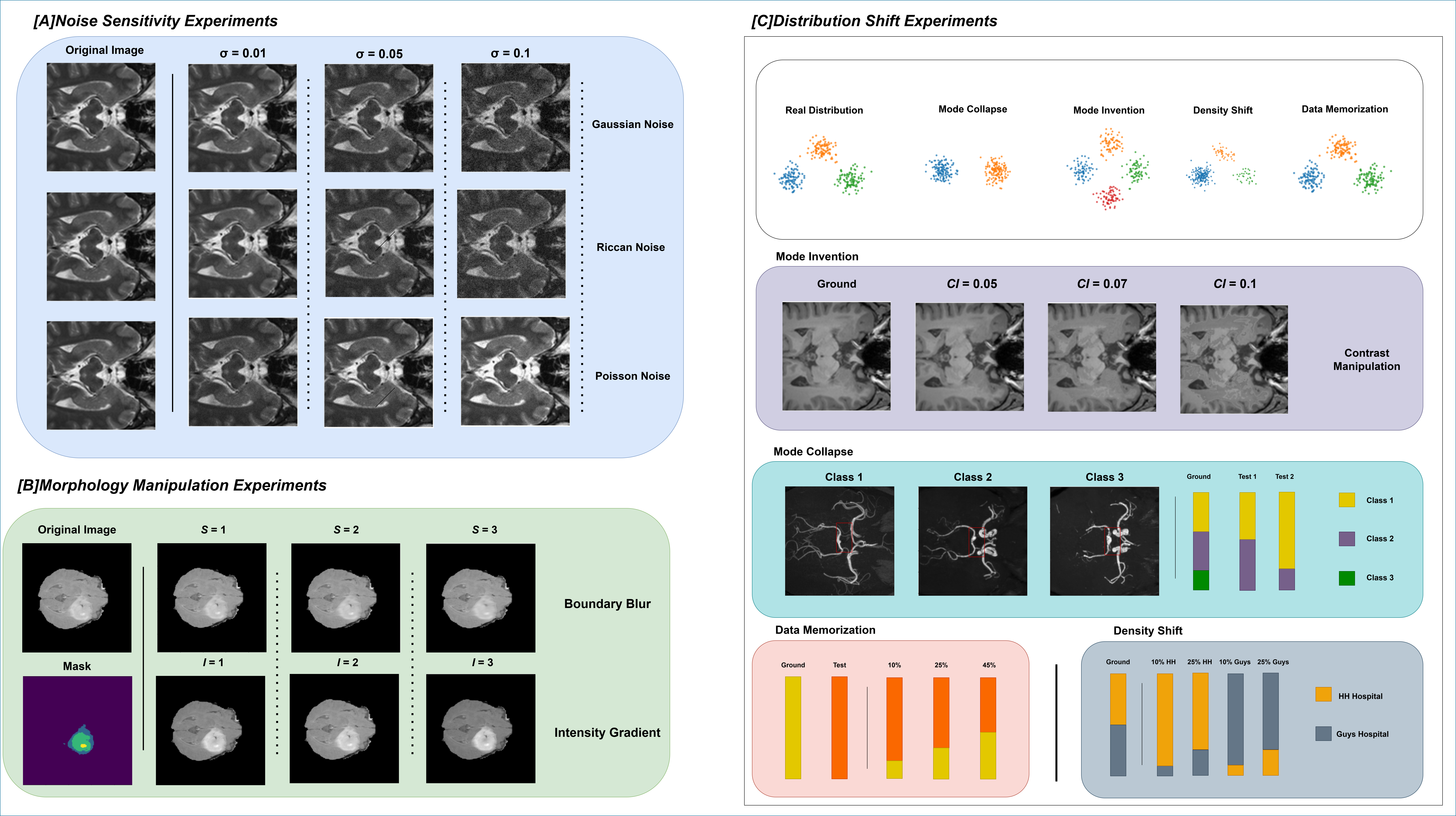}}
\caption{Overview of the controlled perturbation experiments in Phase 1. We apply noise, morphological, and domain-shift manipulations to real medical images to evaluate the sensitivity of NRIQMs.}
\label{fig4}
\end{figure*}

These distribution shift experiments provide a comprehensive evaluation framework. Fig.~\ref{fig4} visually summarises the different controlled perturbation experiments conducted in this phase.

\label{method}

\subsection{Phase 2: Generative Model Performance Evaluation}
\label{sec:phase2_gen_model_eval}

Building upon the insights from the controlled perturbation experiments, this second phase focuses on evaluating synthetic images produced by established generative model architectures. The goal is to assess how well the upstream NRIQMs and the downstream segmentation task performance align when applied to realistic outputs from different types of generative models operating on complex medical imaging data. We evaluate synthetic images corresponding to the target domains of the IXI (vascular MRA) and BraTS (brain tumour MRI) datasets.

\label{subsec:gen_model_setup}

For the generative model evaluation phase, we utilised publicly available, \textbf{pre-trained} implementations representing three prominent generative architectures. The specific models used for each dataset were:
\begin{itemize}
    \item \textbf{For the IXI dataset:} Pre-trained model weights for all three architectures were sourced directly from the study by Deo et al.~\cite{deo2023shapeguidedconditionallatentdiffusion}. These specific implementations were:
        \begin{itemize}
            \item \textbf{VAE:} A Vector Quantised VAE (VQ-VAE)~\cite{VQ-VAE}.
            \item \textbf{GAN:} A Wasserstein GAN (WGAN)~\cite{Wassgan}.
            \item \textbf{DDPM:} The Latent Diffusion Model (LDM) as implemented and detailed in the source paper~\cite{deo2023shapeguidedconditionallatentdiffusion}.
        \end{itemize}
    \item \textbf{For the BraTS dataset:}
        \begin{itemize}
            \item \textbf{GAN:} A pre-trained Wasserstein GAN (WGAN) obtained from the Medigan library/toolkit~\cite{medigan}. The underlying architecture is based on~\cite{Wassgan}.
            \item \textbf{DDPM:} The Latent Diffusion Model (LDM) implementation described by Rombach et al.~\cite{latdiffm}.
        \end{itemize}
\end{itemize}

The set of synthetic images generated by each of these pre-trained models constitutes our \emph{generated image set} for evaluation in this phase. We apply the full suite of upstream NRIQMs to these generated images, comparing their feature distributions against the corresponding real images (designated as the \emph{reference distribution set} from Phase 1). Concurrently, we apply the pre-trained segmentation model (detailed in Section ~\ref{subsec:downstream technique}) to these same generated images to assess downstream task performance.

By systematically applying both upstream quantitative metrics and downstream task validation across controlled perturbations (Phase 1) and generative model outputs (Phase 2), this methodology enables a critical assessment of the NRIQMs. This comparative approach is designed to expose the conditions under which specific metrics succeed or fail, highlighting discrepancies between quantitative scores and practical task performance. Ultimately, this work aims to inform the selection and development of more reliable validation strategies for generative models in medical imaging, paving the way for evaluation frameworks that demonstrably correlate with clinical relevance and robustly identify model failure modes impacting potential downstream applications.

\section{Results}

In this section, we report the findings from our multifaceted evaluation framework, analysing the behaviour of NRIQMs introduced in Section~\ref{sec:metrics}. We present results from both the controlled perturbation experiments (Phase 1) and the evaluation of actual generative model outputs (Phase 2). Complete numerical results for all upstream metrics are provided in the supplementary materials (Section A.1). For the controlled experiments, we compare metric scores obtained from perturbed image sets against baseline scores from real data. Upstream metric sensitivities are often using heat-maps of standardised Z-scores for intuitive comparison (as shown in Figs.~\ref{heat_N}-\ref{heat_Di}). Downstream task performance is assessed quantitatively using the Evanyseg score described in Sec.~\ref{subsubsec:ref_challenge}.

\subsection{Phase 1: Controlled Perturbation Experiments} 

\subsubsection{Noise Sensitivity Results} 

\begin{figure*} 
\centerline{\includegraphics[width=\textwidth]{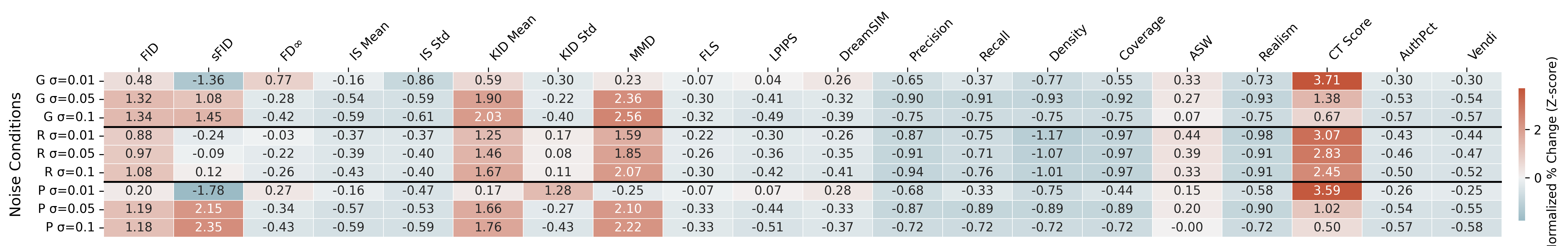}}
\caption{Heatmap illustrating the sensitivity of each evaluation metric to various forms of noise perturbation.}
\label{heat_N}
\end{figure*}

Our initial experiments investigated metric responses to controlled noise injections (Gaussian, Poisson, Rician). The sensitivity of upstream metrics is visualised in Fig.~\ref{heat_N}. As observed, distance-based metrics such as FID, sFID, KID, MMD, and FLS generally demonstrated increased dissimilarity (worsening scores) with rising noise levels, reacting particularly strongly to Gaussian and Rician noise. Learned perceptual metrics (LPIPS, DreamSIM) also consistently captured degradation across noise types. Conversely, several PRDC metrics (Precision, Recall, Density, Coverage) and the Realism Score, while sensitive, often saturated towards worst-case scores even at moderate noise levels (especially Gaussian/Rician), potentially limiting their utility for differentiating finer degrees of degradation. Inception Score exhibited inconsistent trends across noise types: its mean value tended to decrease (worsen) with Gaussian/Rician noise but increase (improve) with Poisson noise, while its standard deviation showed the opposite pattern (improving stability with Gaussian/Rician but worsening with Poisson noise). The Vendi Score remained largely insensitive.

Downstream segmentation performance, assessed via the Evanyseg score, showed varied sensitivity depending on the noise type (Table~\ref{tab:evany_noise}). Gaussian noise injection resulted in only a moderate decrease in segmentation quality compared to the baseline (0.907). Interestingly, Poisson noise, even at the highest level tested, caused minimal degradation in the downstream task score. Rician noise, however, induced a substantial drop in segmentation performance, particularly at higher sigma values, aligning with its physical relevance in MRI and suggesting the downstream task is quite sensitive to this specific noise type. Comparing these findings, while upstream metrics such as FID reacted strongly to both Gaussian and Rician noise, the downstream task was significantly more impacted by Rician noise, highlighting a potential divergence where upstream metric sensitivity doesn't perfectly predict downstream task robustness for all noise types.

\begin{table}[h] 
\centering
\caption{Downstream Evanyseg scores for Noise Sensitivity experiments (Baseline: 0.907).}
\label{tab:evany_noise}
\begin{tabular}{@{}lc@{}}
\toprule
Noise Condition & Evanyseg Score \\
\midrule
\textit{Gaussian} & \\
\quad $\sigma=0.01$ & 0.804 \\
\quad $\sigma=0.05$ & 0.7886 \\
\quad $\sigma=0.1$ & 0.7994 \\
\midrule
\textit{Rician} & \\
\quad $\sigma=0.01$ & 0.904 \\
\quad $\sigma=0.05$ & 0.748 \\
\quad $\sigma=0.1$ & 0.451 \\
\midrule
\textit{Poisson} & \\
\quad $\sigma=0.01$ & 0.893 \\
\quad $\sigma=0.05$ & 0.889 \\
\quad $\sigma=0.1$ & 0.877 \\
\bottomrule
\end{tabular}
\end{table}

\subsubsection{Morphological Manipulation Results} 
\label{sec:morph_results} 

\begin{figure*}[h] 
  \centerline{\includegraphics[width=\textwidth]{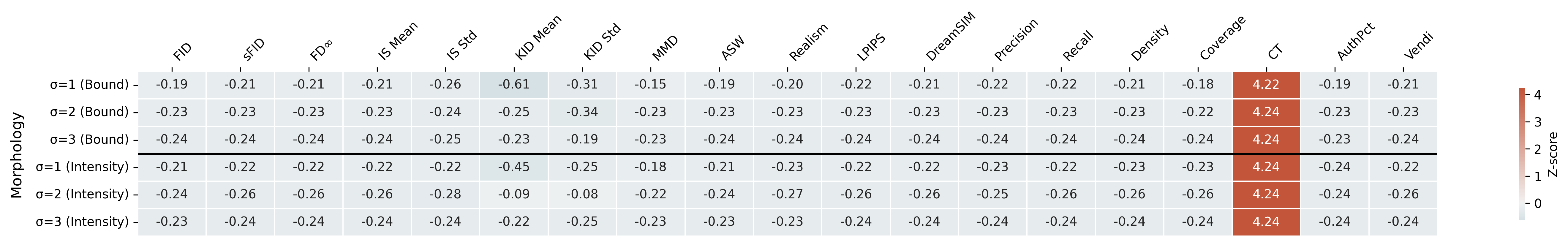}}
  \caption{Heatmap illustrating the sensitivity of upstream evaluation metrics to morphological perturbations. Note the widespread lack of response across most metrics, contrasting with the downstream sensitivity shown in Table~\ref{tab:evany_morph}.} 
  \label{heat_M}
\end{figure*}

Experiments introducing localised morphological manipulations within clinical regions (specifically tumour boundary blurring and internal intensity gradient alterations in BRaTS images) yielded slightly contrasting results between upstream metrics and downstream task performance. As visualised in Fig.~\ref{heat_M}, a profound insensitivity was observed across nearly all tested upstream NRIQMs — including distance-based (FID, sFID, KID, MMD), perceptual (LPIPS, DreamSIM), PRDC metrics, and others such as AuthPct and Realism Score. These metrics showed negligible changes from baseline values even with increasing levels of boundary blurring or internal intensity variation, with only the CT Score exhibiting a minimal reaction.

In contrast, downstream segmentation performance, assessed via the Evanyseg score (Table~\ref{tab:evany_morph}), did show sensitivity to these morphological changes. Both tumour boundary blurring and the introduction of radial intensity gradients caused a measurable decrease in the segmentation quality score, with the degradation generally increasing as the manipulation strength ($\sigma$) increased. While the absolute changes in the Evanyseg score might not be dramatic, the consistent downward trend indicates that the downstream segmentation task was negatively impacted by these subtle alterations to tumour morphology.

This comparison highlights a critical finding: while the downstream task registered a response to these clinically relevant morphological manipulations, the vast majority of upstream NRIQMs completely failed to detect these changes. This underscores the potential unreliability of relying solely on upstream metrics for evaluating the anatomical fidelity and fine-grained structural accuracy of generated medical images.

\begin{table}[h] 
\centering
\caption{Downstream Evanyseg scores for Morphological Manipulation experiments (Baseline: 0.948).} 
\label{tab:evany_morph}
\begin{tabular}{@{}lc@{}}
\toprule
Manipulation & Evanyseg Score \\
\midrule
\textit{Intensity Gradient}& \\
\quad $\sigma=1$ & 0.934 \\
\quad $\sigma=2$ & 0.902 \\
\quad $\sigma=3$ & 0.881 \\
\midrule
\textit{Boundary Blur} & \\
\quad $\sigma=1$ & 0.932 \\
\quad $\sigma=2$ & 0.893 \\
\quad $\sigma=3$ & 0.845 \\
\bottomrule
\end{tabular}
\end{table}

\subsubsection{Data Memorisation Results}
\label{sec:dup_results}

\begin{figure*}[h]
  \centerline{\includegraphics[width=\textwidth]{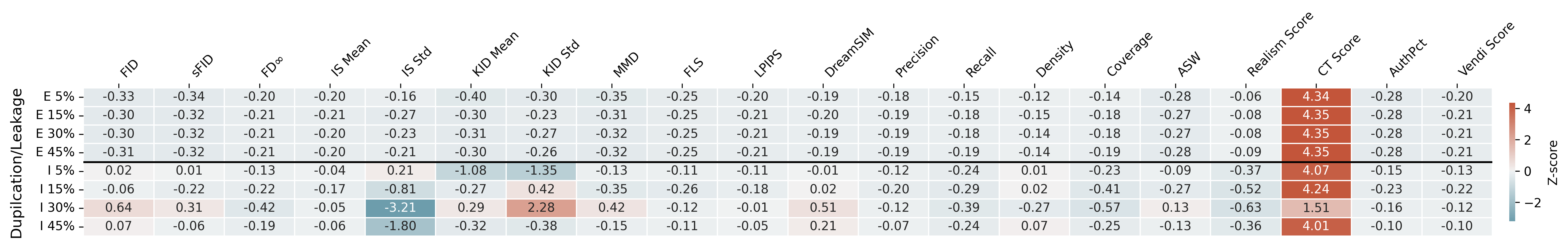}}
  \caption{Heatmap illustrating the sensitivity of upstream evaluation metrics to increasing levels of data memorisation (External: Training set leakage; Internal: Within-set duplication).}
  \label{heat_Dp}
\end{figure*}

Analysis of data memorisation effects, visualised via upstream metric sensitivities in Fig.~\ref{heat_Dp}, highlights more potential assessment pitfalls. As observed, metrics sensitive to distribution overlap (FID, sFID, KID, MMD) showed apparent improvement(lower distance scores) as the level of external duplication (training data leakage) increased. While reflecting increased feature similarity, this trend could be highly misleading if interpreted as better generative model quality in a real scenario. Concurrent with this, AuthPct dropped significantly, correctly identifying the decreased authenticity due to leakage, suggesting that combining distance metrics with AuthPct offers a potential signature for diagnosing external duplication. Conversely, most upstream metrics, including distance-based ones and even diversity metrics such as Vendi Score, showed minimal sensitivity to internal duplication (within-set repetition).

Downstream segmentation performance (Table~\ref{tab:evany_dup}) remained remarkably stable and very close to the baseline score (0.907) across all levels of both external and internal data duplication. This suggests that, for this specific downstream task and evaluation method, performance was largely unaffected by the presence of duplicated images in perturbed image sets, whether leaked from training or repeated internally. This contrasts sharply with the significant shifts observed in upstream metrics such as FID and AuthPct under external duplication, emphasising that upstream distribution similarity does not always predict downstream task robustness or behaviour.

\begin{table}[h] 
\centering
\caption{Downstream Evanyseg scores for Data Memorisation experiments (Baseline: 0.907).}
\label{tab:evany_dup}
\begin{tabular}{@{}lc@{}}
\toprule
Condition & Evanyseg Score \\
\midrule
\textit{External Duplication} & \\
\quad 5\% & 0.908 \\
\quad 15\% & 0.908 \\
\quad 30\% & 0.907 \\
\quad 45\% & 0.909 \\
\midrule
\textit{Internal Duplication} & \\
\quad 5\% & 0.907 \\
\quad 15\% & 0.908 \\
\quad 30\% & 0.912 \\
\quad 45\% & 0.912 \\
\bottomrule
\end{tabular}
\end{table}

\subsubsection{Density Shift, Mode Collapse and Mode Invention Results}
\label{sec:dist_inv_results}

\begin{figure*}[!h] 
  \centerline{\includegraphics[width=\textwidth]{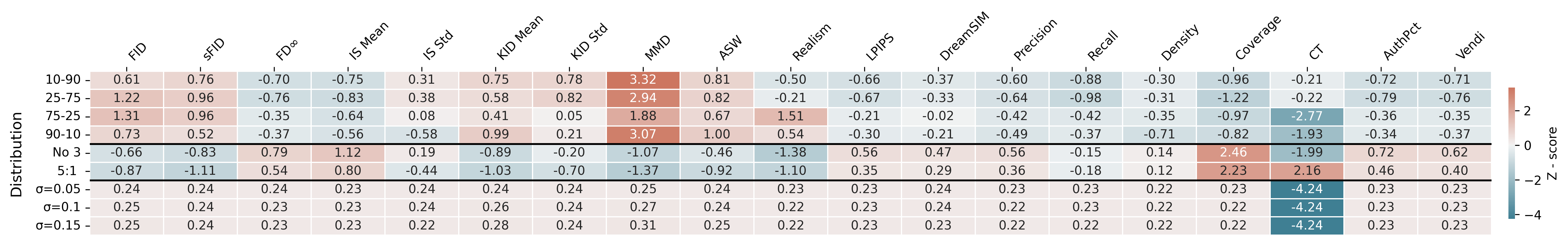}}
  \caption{Heatmap illustrating the sensitivity of upstream evaluation metrics to density shift, mode collapse and mode invention.}
  \label{heat_Di}
\end{figure*}

Experiments altering data distribution (simulating density shifts between scanners and mode collapse by manipulating class proportions) and simulating mode invention (via artificial contrast manipulation) revealed varied sensitivities among upstream metrics (Fig.~\ref{heat_Di}). Distance metrics (FID, sFID, KID, MMD, ASW) generally reacted to these shifts and manipulations, detecting deviations from the baseline distribution. However, their behaviour under mode collapse was potentially misleading, showing score 'improvements' when rare classes were removed. Recall and potentially Realism Score appeared more indicative of such mode coverage failures. AuthPct again showed contrasting diagnostic behaviour, decreasing with external data duplication, but increasing with mode invention, suggesting divergence from the real data manifold. LPIPS remained relatively insensitive to density shifts and mode collapse but reacted moderately to mode invention, while DreamSIM reacted counter-intuitively to mode invention. FD$\infty$ and Vendi Score continued to show limited sensitivity across these diverse challenges.

Remarkably, downstream segmentation performance again remained almost entirely unaffected by density shifts, as well as by mode collapse and invention (Table~\ref{tab:evany_dist_inv}). The Evanyseg scores stayed very close to the respective baselines across all tested conditions, including varying scanner proportions (density shift), removal or imbalance of anatomical classes (mode collapse), and artificial contrast invention (mode invention). This striking insensitivity of the downstream task performance contrasts significantly with the responses observed in many upstream metrics, particularly the distance-based ones and AuthPct, which clearly detected these distributional manipulations. This divergence strongly suggests that detecting a statistically significant shift in upstream feature distributions does not necessarily imply a meaningful impact on the utility of the images for this specific downstream segmentation task.

\begin{table}[h] 
\centering
\caption{Downstream Evanyseg scores for Distribution Shift and Mode Invention experiments.}
\label{tab:evany_dist_inv}
\begin{tabular}{@{}lc@{}}
\toprule
Condition & Evanyseg Score \\
\midrule
\textit{Density Shift (Scanner Proportions)} & \\
\quad Baseline (50-50) & 0.907 \\ 
\quad 10-90 & 0.904 \\
\quad 25-75 & 0.909 \\
\quad 75-25 & 0.902 \\
\quad 90-10 & 0.901 \\
\midrule
\textit{Mode Collapse (PCoA Classes)} & \\
\quad Baseline (All Classes) & 0.907 \\ 
\quad No Class 3 (50-50) & 0.904 \\
\quad No Class 3 (5:1 Imbalance) & 0.901 \\
\midrule
\textit{Mode Invention (Contrast Manip.)} & \\
\quad Baseline & 0.907 \\ 
\quad $\sigma=0.05$ & 0.909 \\
\quad $\sigma=0.07$ & 0.909 \\ 
\quad $\sigma=0.1$ & 0.908 \\ 
\bottomrule
\end{tabular}
\end{table}

\subsection{Phase 2: Generative Model Performance Evaluation Results}
\label{sec:gen_model_results} 

\begin{figure*}[h] 
  \centerline{\includegraphics[width=\textwidth]{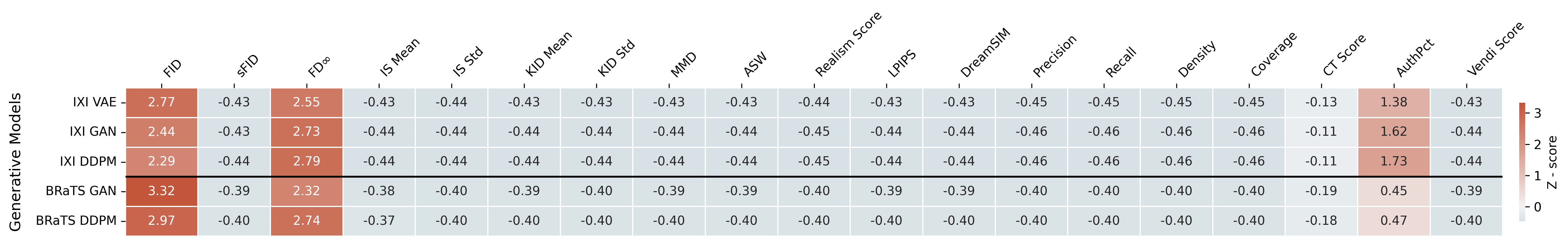}}
  \caption{Heatmap illustrating the sensitivity of upstream evaluation metrics across different generative model outputs (VAE, GAN, DDPM on IXI and BRaTS datasets). Scores represent standardised differences relative to a baseline or between models, highlighting how metrics differentiate model outputs.} 
  \label{heat_G}
\end{figure*}

Moving beyond controlled perturbations, we evaluated synthetic images produced by representative pre-trained generative models — a VAE, a GAN, and a DDPM — on the IXI and BRaTS datasets. We applied both the upstream NRIQMs (detailed in Sec.~\ref{sec:metrics}) and the downstream segmentation task evaluation (using the Evanyseg method detailed in Sec.~\ref{subsubsec:ref_challenge}), with results presented in Fig.~\ref{heat_G} and Table~\ref{tab:evany} respectively to enable direct comparison of these different evaluation modalities on generative model outputs.

The upstream metric results reveal a striking pattern: nearly all evaluated NRIQMs indicated substantial deviations between the generated images (regardless of model type) and the real data baseline distributions. As shown in Fig.~\ref{heat_G}, distance-based metrics (FID, sFID, KID, MMD, ASW) yielded markedly high dissimilarity scores for VAE, GAN, and DDPM compared to real data. While minor differences existed between models (e.g., DDPM often slightly better than GAN/VAE on some distance metrics for IXI), the overall magnitude of the deviation from the real baseline was typically large for all architectures, making nuanced quality ranking based solely on these metrics challenging. Furthermore, PRDC metrics generally collapsed to near-zero values, and AuthPct saturated at 100\%, confirming that although these metrics effectively distinguished all generated distributions from the real one, they provided little discrimination between the generative models themselves. LPIPS showed moderate degradation across models, DreamSIM exhibited potentially counter-intuitive increases, and IS varied (notably high for DDPM on BRaTS). Overall, the upstream metrics strongly signalled a distributional mismatch for all models but struggled to differentiate them in a way that reflected expected visual quality differences.

In contrast, evaluating the models via the downstream segmentation task using Evanyseg scores, as shown in Table~\ref{tab:evany}, provided a distinctly different and more discriminative assessment, aligning better with anticipated visual fidelity. On the IXI vessel segmentation task, the DDPM achieved the highest score (0.91), significantly outperforming the GAN (0.67) and also surpassing the VAE (0.86). Similarly, for BRaTS tumour segmentation, the DDPM (0.78) demonstrated clearly superior performance compared to the GAN (0.56). This downstream ranking reflects the typical expectation that DDPMs often produce images with higher structural coherence suitable for such tasks compared to VAEs or standard GANs. Intriguingly, 
results on IXI dataset highlight a direct contradiction: while the GAN produced 'better' upstream scores than the VAE on several distance metrics (visible in Fig.~\ref{heat_G} data), its output was significantly less suitable for the downstream segmentation task compared to the VAE's output (0.67 vs 0.86). This discrepancy starkly underscores the potential unreliability of using only upstream metrics for model selection or quality assessment.

This observed divergence between the generally poor and undiscriminating upstream metric scores and the more differentiated downstream task performance reinforces a critical limitation of relying solely on many common distribution-focused NRIQMs. While these upstream metrics can effectively detect the \emph{presence} of a distributional shift between generated and real images, they appear inadequate in characterising the \emph{nature} of this shift in a way that consistently aligns with downstream task suitability across different generative architectures. These results strongly advocate for incorporating downstream task evaluations as a more meaningful and reliable method for assessing the practical utility of generative models in medical imaging applications.

\begin{table}[h] 
\caption{Calculated Evanyseg Dice scores for downstream evaluation of generative models.} 
\label{tab:evany}
\begin{tabular}{|p{2cm}|p{2cm}|p{2cm}|}
\hline
\textbf{Model}     & \textbf{IXI Vessel Segmentation} & \textbf{Brats Tumour Segmentation}\\ \hline
\textbf{VAE}      & 0.86               & -                 \\ \hline
\textbf{GAN}      & 0.67               & 0.56               \\ \hline
\textbf{DDPM}      & 0.91               & 0.78               \\ \hline
\end{tabular}
\end{table}

\begin{table*}[htbp]
\centering
\caption{This table summarises some of the key findings and Recommendations for Evaluating Generative Medical Images with NRIQMs}
\label{tab:summary_takeaways_refined}

\begin{tabularx}{\textwidth}{@{} >{\centering\arraybackslash}p{0.5cm} >{\raggedright\arraybackslash}X >{\raggedright\arraybackslash}p{4.5cm} >{\raggedright\arraybackslash}X @{}} 
\toprule
\textbf{\#} & \textbf{Recommendation} & \textbf{Context} & \textbf{Implication / Why it Matters for Reliable Evaluation} \\
\midrule
1 & \textbf{Prioritise Task-Based (Downstream) Evaluation:} It proved more reliable and discriminative for assessing practical model utility than upstream metrics alone. 
  & Upstream vs. Downstream Comparison (All Models); Evanyseg results vs. Upstream metrics, e.g. FID/KID etc.
  & Task performance better reflects clinical relevance, effectively differentiates between models (e.g., DDPM vs. GAN), and aligns more closely with desired visual/functional quality. \\

\addlinespace

2 & \textbf{Use Distance Metrics With Extreme Caution:} These metrics are highly prone to misleading scores despite sensitivity to global data distribution shifts.
  & FID, KID, MMD, ASW, sFID
  & Scores can falsely suggest improvement with data leakag/memorization or mode collapse. Correlated poorly with task performance across models. Unreliable for quality ranking alone. \\

\addlinespace

3 & \textbf{Recognise Metric Blindness to Local Detail:} Most tested upstream metrics fail to detect fine-grained anatomical or pathological errors.
  & All Upstream Metrics remained largely unchanged in Local Boundary\&Intensity Manipulation experiments.
  & Critical Gap: Cannot ensure local anatomical fidelity, potentially missing subtle but clinically significant errors. Poses a safety concern if used as primary validation. \\

\addlinespace

4 & \textbf{Actively Investigate Mode Collapse \& invention:} Do not rely solely on distance metrics, which can be deceptively optimistic.
  & FID/KID (misleading improvement), Recall potentially signals degradation, Downstream testing insensitive.
  & Models might fail to generate rare but critical conditions, necessitating targeted evaluation beyond global scores, e.g. monitoring recall and performing visual \& task-specific inspection. \\

\addlinespace

5 & \textbf{Diagnose Data Memorisation Using Combined Signals:} Look beyond seemingly 'improved' distance scores.
  & FID/KID (misleading improvement), AuthPct (authenticity score drop).
  & Combining these signals helps differentiate genuine model performance from artificial score inflation caused by training data repetition in the test set. \\

\addlinespace

6 & \textbf{Avoid Metrics Demonstrated to be Unreliable or Insensitive:} Certain metrics consistently provided little useful information or behaved erratically.
  & Vendi Score, FD$\infty$ (Insensitive); IS, DreamSIM (Inconsistent/Counter-intuitive).
  & These specific metrics were unsuitable or unreliable for the challenges tested in these controlled perturbation experiments\\

\addlinespace

8 & \textbf{Adopt a Multifaceted Evaluation Strategy:} No single metric provides a complete or trustworthy picture of generative model quality.
  & Synthesis of all experimental findings.
  & Robust assessment requires combining cautiously interpreted upstream metrics, essential downstream task validation, and potentially expert visual inspection. \\

\bottomrule
\end{tabularx}
\end{table*}

\section{Discussion}
\label{sec:discussion}

Our systematic evaluation investigated the reliability of commonly used NRIQMs for assessing generative models in medical imaging, comparing quantitative upstream metric scores against downstream task performance. Multidisciplinary perspectives on simulation, data, and model evaluation help to contextualise our findings, which reveal significant limitations in relying solely on upstream metrics and emphasise the need for purpose-driven validation frameworks~\cite{BokulichParker2021, Parker2023_Eval}. Generated images, unlike direct clinical measurements, are outputs of complex computational processes – akin to simulations – whose epistemic status is inherently indirect~\cite{Parker2021_Virtually, Parker2022_Evidence}. They represent the model's learned approximation of reality, potentially lacking fidelity due to uncaptured factors or 'unknown unknowns' inherent in the real-world data generating process, placing them epistemically 'one step behind' direct experimentation or measurement~\cite{Roush2018}.

A primary empirical observation was the sensitivity of distance-based metrics (e.g., FID, KID) to global distributional characteristics, yet their potential to mislead. Their 'improvement' with data duplication or mode collapse highlights a failure to capture crucial aspects of generative quality beyond statistical moments. Furthermore, their poor correlation with downstream task performance across different model architectures (VAE, GAN, DDPM) suggests they often fail the test of 'adequacy-for-purpose'~\cite{BokulichParker2021, Parker2023_Eval}. The hypothesis that the homogeneity of medical images compared to diverse natural scenes might challenge these metrics, especially those using features from natural image models, warrants further investigation. If metrics cannot reliably distinguish subtle yet critical pathological variations, their utility diminishes significantly.

Other metrics showed varied utility. The saturation of PRDC metrics limited their discriminative power. AuthPct and Recall, however, showed promise as diagnostic tools for specific failure modes (leakage, mode invention, mode collapse), aligning with the idea that data evaluation involves targeted assessment for specific purposes~\cite{BokulichParker2021} rather than seeking a single quality score. The consistent insensitivity of Vendi Score and FD$\infty$ underscores the difficulty of finding universally effective NRIQMs.

Perhaps the most critical finding, viewed through the lens of distinguishing data from models~\cite{Leonelli2019} and experiment from simulation~\cite{Roush2018}, is the profound insensitivity of almost all upstream metrics to localised morphological manipulations. These metrics, treating generated images as data points in a feature space, failed to detect alterations in boundary sharpness or internal gradients – features crucial for clinical interpretation and downstream tasks such as segmentation. While our downstream evaluation did register performance degradation (although limited) under these conditions, the upstream metrics largely failed. This highlights a significant risk: generative models might produce anatomically flawed images that pass common quantitative metrics checks but would be unsuitable - or even hazardous - for applications such as augmenting training datasets for developing clinical decision support tools, potentially leading these tools to learn incorrect anatomical representations. These finding highlight that upstream metrics often lack the representational granularity needed to validate anatomical fidelity. This aligns with the philosophical view that evaluating representations requires considering the specific purpose; metrics optimised for global distribution matching may be inadequate for purposes requiring local structural accuracy~\cite{BokulichParker2021}.

The divergence between upstream and downstream results, particularly the VAE/GAN contradiction on IXI, strongly supports the 'adequacy-for-purpose' approach to evaluation~\cite{Parker2023_Eval, BokulichParker2021}. It suggests that synthetic data which misrepresent the real distribution according to upstream metrics can still be adequate for specific downstream tasks. However, our finding that the downstream task was insensitive to various distributional shifts (data memorisation, density shifts, mode invention) that were detected by upstream metrics serves as a crucial counterpoint. It implies that downstream evaluation, while necessary for assessing utility, is not sufficient alone. It evaluates fitness for one purpose and may overlook other significant flaws, aligning with Parker's discussion of model evaluation challenges~\cite{Parker2023_Eval}.

These findings carry significant implications for deploying generative models in safety-critical medical applications. Relying on easily computed upstream metrics - without validating downstream utility - risks deploying flawed models or clinical tools trained on inadequate synthetic data. Safety is highly context-dependent, and even slight variation in the clinical settings can undermine the validity of the existing evidence ~\cite{habli2025big}. Further, viewing generated images as akin to simulation outputs, offering potentially valuable insights~\cite{Parker2022_Evidence}, demands a rigorous validation process focused on their fitness for specific clinical purposes~\cite{Parker2023_Eval}. Therefore, we advocate strongly for a multifaceted evaluation framework, integrating downstream task validation with context-aware interpretation of appropriate upstream metrics to ensure both utility and safety.

\textbf{Limitations of this Study:} Our findings are primarily based on two specific MRI datasets and one downstream task (segmentation). Metric performance and the upstream-downstream relationship may differ across other modalities, anatomical regions, or clinical tasks. The specific generative models and perturbation methods used may not cover all possibilities. Furthermore, the downstream evaluation relied on a specific pre-trained segmentor and the Evanyseg method, introducing dependencies on these components.

\textbf{Suggestions for Future Research:}  Our findings underscore several critical areas for future research, aligning with recognised needs in the evaluation of generative medical imaging. Firstly, addressing the \textbf{limitations of existing metrics}, particularly the noted issues with widely-used distance metrics such as FID in medical contexts~\cite{abdusalomov2023evaluating} and the general insensitivity of upstream metrics to localised anatomical details observed in our work, necessitates the development of novel metrics. Future metrics should aim for enhanced sensitivity to fine-grained, clinically relevant features and validation against specific medical image failure modes. Secondly, the significant divergence we observed between upstream metric scores and downstream task performance reinforces the growing consensus on the importance of downstream task evaluation~\cite{jha2022objective}. Further work is needed to validate findings across a broader spectrum of clinical tasks and imaging modalities, and to develop robust, potentially reference-free, methods for assessing downstream utility reliably, possibly incorporating insights from approaches using human feedback\cite{pak2023aligning}. Integrating insights from both improved upstream metrics and rigorous downstream task validation will be key to building truly trustworthy generative models for healthcare. An integrated approach, grounded in a purpose-driven evaluation philosophy~\cite{BokulichParker2021, Parker2023_Eval}, is vital for the responsible progress of generative models in healthcare.

\section{Conclusion}
\label{sec:conclusion}

Our comprehensive assessment revealed significant limitations in the utility of many commonly used NRIQMs for evaluating generative models in medical imaging. While sensitive to certain global distributional characteristics, these metrics frequently correlate poorly with practical utility, often fail to detect clinically relevant localised inaccuracies, and can provide misleading assessments regarding issues such as mode collapse or data memorisation. We observed a stark divergence between upstream metric scores and downstream task performance, with the latter often proving more discriminative but also exhibiting its own limitations, such as insensitivity to certain distributional shifts. Therefore, we conclude that neither upstream nor downstream evaluation alone is sufficient for reliable validation. Robust and trustworthy assessment of generative models necessitates a multifaceted framework, integrating insights from relevant downstream task evaluations with the cautious interpretation of appropriately selected upstream metrics. Addressing the identified gaps requires future research focused on developing novel metrics with enhanced sensitivity to localised anatomical features and clinical relevance. This should be accompanied by broader validation across diverse medical tasks and modalities, as well as continued refinement of robust downstream task evaluation methodologies. Moving forward with such integrated and improved evaluation strategies is essential for the safe and effective translation of generative models into clinical practice.

\section*{Acknowledgements}
This work was supported by the Centre for Assuring Autonomy, a partnership between Lloyd’s Register Foundation and the University of York. AFF acknowledges support from the Royal Academy of Engineering under the RAEng Chair in Emerging Technologies (INSILEX CiET1919/19), ERC Advanced Grant – UKRI Frontier Research Guarantee (INSILICO EP/Y030494/1), the UK Centre of Excellence on in-silico Regulatory Science and Innovation (UK CEiRSI) (10139527), the National Institute for Health and Care Research (NIHR) Manchester Biomedical Research Centre (BRC) (NIHR203308), the BHF Manchester Centre of Research Excellence (RE/24/130017), and the CRUK RadNet Manchester (C1994/A28701).

\section*{Competing interest}

The authors declare no competing interests.

\bibliographystyle{plain}
\bibliography{refs}

\end{document}